# Lessons for GenAI Literacy from a Field Study of Human-GenAI Augmentation in the Workplace

Aditya Johri
*Information Sciences & Technology*
George Mason University
Fairfax, VA, USA
johri@gmu.edu

Johannes Schleiss
*AI Lab*
Otto-von-Guericke University
Magdeburg, Germany
johannes.schleiss@ovgu.de

Nupoor Ranade
*Rhetoric & Technical Communication*
Carnegie Mellon University
Pittsburgh, PA, USA
nupoor@cmu.edu

*Abstract*—Generative artificial intelligence (GenAI) is increasingly becoming a part of work practices across the technology industry and being used across a range of industries. This has necessitated the need to better understand how GenAI is being used by professionals in the field so that we can better prepare students for the workforce. An improved understanding of the use of GenAI in practice can help provide guidance on the design of GenAI literacy efforts including how to integrate it within courses and curriculum, what aspects of GenAI to teach, and even how to teach it. This paper presents a field study that compares the use of GenAI across three different functions - product development, software engineering, and digital content creation - to identify how GenAI is currently being used in the industry. This study takes a human augmentation approach with a focus on human cognition and addresses three research questions: how is GenAI augmenting work practices; what knowledge is important and how are workers learning; and what are the implications for training the future workforce. Findings show a wide variance in the use of GenAI and in the level of computing knowledge of users. In some industries GenAI is being used in a highly technical manner with the deployment of fine-tuned models across domains. Whereas in others, only off-the-shelf applications are being used for generating content. This means that the need for what to know about GenAI varies, and so does the background knowledge needed to utilize it. For the purposes of teaching and learning, our findings indicated that different levels of GenAI understanding need to be integrated into courses. From a faculty perspective, the work has implications for training faculty so that they are aware of the advances and how students are possibly, as early adopters, already using GenAI to augment their learning practices.

## I. INTRODUCTION

Educating the future technology workforce, including engineering and computing students, requires building a more robust understanding of the impact of artificial intelligence (AI), and especially generative AI (GenAI) on how technological work gets done [1]. Microsoft and LinkedIn's 2024 Work Trend Index Annual report [2] released in May 2024 reports that use of generative AI has nearly doubled in the last six months, with 75% percentage of global knowledge workers using it. It also reported that if an organization did not have guidance on the use of GenAI, employees were taking things into their own hands and hiding their use of GenAI. They reported that 78% of GenAI users, in their sample, brought their own applications to work. Furthermore, they found that 53% of respondents who use GenAI at work worried that using it on important work tasks made them look replaceable.

From the organization perspective, according to the report [2], a majority (55%) of leaders said they were concerned about having enough talent to fill roles in the year ahead as use of GenAI increased. Furthermore, 66% of leaders say they would not hire someone without AI skills and 71% say they would rather hire a less experienced candidate with AI skills than a more experienced candidate without them. This means that junior candidates who have AI skills may have an edge as 77% of leaders expressed a preference for giving early-in-career talent with AI skills greater responsibilities. Finally, the report provides a window on how power users of GenAI reoriented their work patterns. The power users are 56% more likely to use AI to catch up on missed meetings, to analyze information (+51%), to design visual content (+49%), to interact with customers (+49%), and to brainstorm or problem-solve (+37%). Furthermore, the power users are moving past efficiency gains in individual tasks and are 66% more likely to redesign their business processes and workflow with GenAI. Overall, the report, even though it is by a private firm with a stake in selling and advancing GenAI applications, firmly establishes the high integration of GenAI in the workplace.

Given the advances in current and possible future use of GenAI in the workplace, we need a better understanding of the scope of where this new technology is used, how effective it is, and what are its limitations in practice, i.e. integration of GenAI within the workplace ecology [3]. This knowledge can then be used to design courses, curriculum, and training to prepare students for the technical competency required as well as professional skills that are essential for being successful in the workforce [4]. With GenAI, it is also important to do this research to move away from the debates on plagiarism and cheating that have dominated the educational landscape and work towards integrating it in teaching and learning in a more productive manner [5]. Consequently, it is important to undertake research studies of how professionals are using AI and GenAI to provide students the capability to work with GenAI, including emerging skills such as prompt engineering [6] [7]. This will also enable us to achieve the overall important goal of improving students' AI literacy [8], [9].

Within the spectrum of AI, GenAI is unique in that it provides affordances not only for those with high expertise in computing to use it, but even those who are technologically literate to a lesser degree can use the conversational interfaces-based applications such as ChatGPT and Dall-e with high proficiency [10]. Therefore, the use of GenAI seemingly has the potential for a much larger impact across a range of jobs and industries. Although there is necessarily an element of hype around GenAI, similar to any potentially transformative technology, there is also increasing evidence of its impact across industry functions. For instance, those in the technical writing, marketing, and consulting industry are integrating the use of GenAI, especially ChatGPT and related LLM-driven applications, across many functions. With the release of multi-modal GenAI application, this is likely to increase across industries such as customer service. There are also case studies documenting the use across a range of jobs that require software development or coding.

Given the relatively new adoption of GenAI in the workplace, there is a lack of consistency in how it is being used but the experimental ways in which people are using it is a good indicator of its capabilities. Therefore, this is an apt time to undertake a preliminary study that can inform future work in the area. With this goal in mind, we conducted a comparative study with project team members in three organizations to examine how GenAI is being used or experimented with across different kinds of projects. We intentionally picked companies and projects that provided us with a spectrum of work from very highly technical to less technical in nature. Through our study we draw implications for the importance of studying GenAI use in both highly technical workplaces but also aligned spaces where the technology is used largely as an end-user facing application. We also reflect on the nature of AI literacy in this context – what it means, how we can develop it better. Overall, we take an augmentation perspective, that is, how is GenAI able to work with humans to make their experiences and the work outcomes better.

## II. THEORETICAL FRAMEWORK

### A. Human-GenAI Augmentation

For millennia, the ability of humans to augment their physical and mental activities with tools has been a defining characteristic of the species. From simple physical tools such as hammers and sickle, to highly sophisticated ones such as language and writing, and then to more advanced ones such as printing press to calculator, the ability of humans to augment their capacity is critical to its achievements. Human augmentation is the field that looks specifically at the abilities of how humans and their functioning can be enhanced or augmented through the use of technology, including physical or medical technology, in addition to newer forms of digital technologies. According to Raisamo et al. [11], in the current context, human augmentation can be subdivided into three categories: 1) *augmented senses* that involves augmenting vision, hearing, haptic sensation, etc., 2) *augmented action* which involves things like motor movement, amplified force, remote presence, etc., and 3) *augmented cognition* which is focused on information-based interaction or adaptation. Although the three forms of augmentation are correlated and inform each other, in this paper, we focus on augmented cognition and how technology shapes cognition and cognitive activities.

With the rise of AI, there is now a serious interest in studying intelligence augmentation [12] to understand the augmentation that comes with the use of AI-driven technologies and applications [13], including robots [14]. This focus on augmentation is driven by the awareness that full automation is far away and also that automation can have severe repercussions for workforce development. Therefore, scholars have argued that in order to enhance human functioning, it is important for AI to complement human skills [15] and work in a symbiotic manner [16]. Consequently, in our research, the human-AI augmentation relationship was one of the core focus areas especially as work is a very diverse activity and the context of where it takes place, and how shapes its outcomes. Furthermore, any context or organization creates its own culture and cultural practices, and to understand augmentation, these aspects have to be taken into account.

### B. Workplace Studies

Studies of professional work is a core theme of research within engineering and computing education [17]–[19] and professional work practices of technology workers have been an area of intense studies for decades [20]. The initial work focused on information systems design and as the nature of work changed significantly with new digital technologies and work became even more entangled with information systems, including applications like email, it became important to understand how a symbiotic workplace could exist. Findings from this work have emphasized that even what appear to be most mundane of tasks within the workplace can be cognitively demanding and that over time workers develop their own ways to accomplishing them, often using technologies in ways they were not designed for. Furthermore, studies related directly to situated cognition and augmentation of cognitive activities through external means [21] emphasize that to understand how technology impacts work, it is important to account for the specifics of how people accomplish tasks, including the tacit aspects of work practice. Humans are cognitively purposeful by design and often come up with their own unique and novel ways of working. They continue to develop skills and knowledge overtime and across domains and functional areas [22].

The use of GenAI in the workplace is very recent but preliminary studies show that in industries, especially those focused on writing and programming, the use of GenAI applications is on the rise. Technical communicators use them for summarizing complex scientific concepts and generating scientific reports [23]. Workers are using it not only for im-proving their writing, but also for brainstorming, streamlining their workflows, increasing efficiency, and developing content [24]. In software development, the integration of GenAI into development environments has led increased it's use GenAI. For instance, user interface researchers are using it to create templates, transcribe audio data, do basic thematic coding, automate aspects of their work such as feedback. It still cannot provide a cultural context or nuance critique or even assist with intricate collaboration [25]. In other words, consistent with prior work on work practices, we are already starting to see how GenAI is finding novel and unique uses in some cases and getting embedded in work tasks. We also see the need though for conducing more nuances studies focused on GenAI use in order to better understand its symbiosis within work.

*C. Human-GenAI Literacy*

The final area of prior work that informs the framework for this research is work on AI awareness and literacy. AI literacy is an important area for engineering and computing students as across domains AI knowledge and skills are becoming critical in the workplace [26]. Already many scholars have started to study AI literacy and also design curricula and training for raising awareness of AI and providing skills, ranging from basic to highly advanced.

Within this, GenAI has not been studied much and it is important to focus on it. We need to be careful that AI is augmenting and not stunting intelligence [27]. Especially when it comes to preparation of the future workforce, we need a more comprehensive understanding to provide different kinds of support necessary to develop the requisite expertise. We also need to be careful as to not overburden students with learning concepts that are not necessarily important or at least not until they can use it fluently at a basic level.

Finally, it is also important to look into how students can be prepared for future learning. Learning is a continuous activity across an engineering career and the change in technologies means that engineers have to be prepare to keep learning to work with technological advancements [22]. Therefore, in addition to specific skills and technical competency, higher-level metacognitive skills are important [28]. GenAI, thus, needs to be incorporated across higher education in a systematic manner [29].

Almatrafi et al. [8] conducted a systematic review of articles on AI literacy including those that conceptualize AI literacy, conduct AI literacy efforts, and develop instruments to assess AI literacy. They also included AI literacy studies across a range of population including adults in the workforce and through a content analysis synthesized six key constructs of AI literacy: Recognize, Know and Understand, Use and Apply, Evaluate, Create, and Navigate Ethically. In this paper we apply this framework across the case studies conducted to identify the prevalence (high, medium, low, none) of each construct across the participants in the studies.

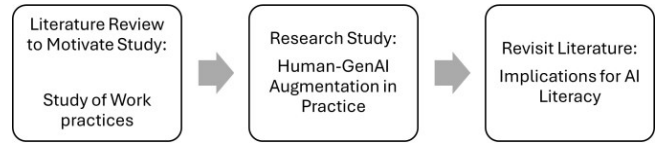

Fig. 1. Research Process

*D. Research Questions*

Overall, based on review of the prior work, and the goals of this research, the following questions were identified:

1. How is GenAI augmenting work practices? a. What specific activities are impacted? b. What level of use is being made?

2. How are users learning to work with GenAI and what knowledge do they need? a. What resources are being used to learn? b. What level of knowledge is needed?

3. What future do participants foresee for future workforce development? a. What are the challenges? b. What needs to be done?

III. RESEARCH STUDY

*A. Research Approach*

Overall, augmentation in the workplace is a complex undertaking and the workplace studies literature recommends that to understand augmentation it is important to focus on "naturally occurring workplace activities" and examine how work is accomplished through in-depth studies that include a focus on the context [30]. Our research started with a literature review to identify areas to investigate as well as the approach to use (see Figure 1). Subsequently, drawing on prior work on professional work practices research, we undertook a qualitative field study using interviews as the primary methodology [31]. This approach advocates for an interpretive understanding of the research context as well as data with the goal of elucidating more situated and contextual insights into work practices [32]. A range of scholars have used this approach to study work practices in the technology sector starting in the early 1990s. This draws on scholarship in sociology, anthropology, and organizational studies [33].

Consistent with a human subjects approach, the study received approval by the Institutional Review Board and consent was taken from all participants. All participants were adults and all interviews were conducted in English and recorded with the permission of the participants.

*B. Research Sites and Participants*

Data was collected with professionals working in three different organizations (refer to Table I for an overview). This purpose sampling was done to get multiple perspectives on how GenAI was being used across projects and work tasks. Data was collected across multiple cities in India.

**Research Site 1: Product development - Project: Concept**: The first organization that was part of the study worked on a range of product development projects for different domains such as finance and human resources. The clients either came to them with an objective or brief or the organization created proof of concepts that it demonstrated to potential clients. The company identified itself as an IT product development enterprise with over 250 professionals and had a portfolio of over 100 products. In terms of specific technologies, it work with the mobile space (Android and iOS), and on Big Data, Machine Learning, Blockchain, AR/VR, and IoT. The company supported entrepreneurs, startup teams, as well as Fortune 500 corporations. The workforce was organized in small as well as large teams, depending on the requirement of the projects.

For the purposes of this study, one project called "Concept" was the focus and the study was conducted with team members working on that project. The primary purpose here was to develop proof-of-concepts of how new and emerging technologies could be used for different domains (e.g., machine learning, deep learning, AI, data science) and use these concepts to demonstrate the ideas to clients. Participants working on the project were highly experienced with software design and development and had high technical competence. One of the participants had a PhD while others had a masters. They had extensive experience with data science, data mining, machine learning, and algorithms.

**Research Site 2: Software development - Project: Code**: This firm identified itself as a software and services firm that specialized in platform application refactoring and migrations. That is, its expertise was in helping clients transition to newer technologies and a lot of the work it did was code migration. It served clients primarily in the area of finance, human resources, healthcare, and telecommunications. The firm employed around 75 employees. The project that was examined for this study "Code" had employed GenAI as part of the development environment to accelerate the coding process and also produce higher quality code. The team consisted of a manager, an architect, and two junior team members.

The expertise and experience of the team working on the project varied. The project manager had extensive experience managing software projects but also as an individual technical contributor. The project architect was a senior team member and technical lead who designed the framework for the project. Two relatively junior software developers with little experience worked as individual contributors to the project. While the two senior people had extensive experience with software development, the junior members were relative novices with only a basic experience with programming and no expertise in the language being used on the project.

**Research Site 3: Content development: Project: Content**: The final research sites was a digital marketing company with about 25 employees and 10 contractors. The company provided services related to advertising campaigns, online and social media, including text, images, and interactive elements. The company also provided search engine optimization (SEO) expertise. and worked across a range of domains or areas. The project "Content" studied here was focused primarily on producing original content for digital campaigns.

TABLE I
INFORMATION ABOUT PARTICIPANTS

| | Project | Role and Responsibility | Exp (yrs) |
|---|---|---|---|
| P1 | Concept | Senior member and team lead for multiple client and internal projects; high technical expertise and contributions to projects. | 15+ |
| P2 | Concept | Senior technical contributor; develops proof-of-concepts projects | 10+ |
| P3 | Concept | Senior technical contributor; proof-of-concepts and client projects | 5+ |
| P4 | Concept | Individual contributor to the project. | 2+ |
| P5 | Concept | Individual contributor to the project. | 2+ |
| P6 | Code | Senior level manager with overall oversight of all aspects of the project, including design and technical contributions; owner of the project discussed. | 20+ |
| P7 | Content | Co-owner and manager; client relationship management. | 15+ |
| P8 | Content | Project lead; responsible for content creation and client relations. | 5+ |
| P9 | Content | Project co-lead; visual content creation. | 3+ |
| P10 | Content | Project contributor; text-based content creation and client relations. | 2+ |

The manager of the team who was also the co-owner of the firm had over two decades of experience in digital marketing. The second senior most team member had around 5 years of experience and led the project. The other team members had 3 and 2 years of experience; one had experience in text and writing while the other on visual design, including images. All team members had high expertise in their domain but the technical fluency and expertise varied. Since they all had undergraduate and graduate degrees in humanities or social sciences, the technical knowledge they had came from work experience and online sources.

*C. Data Collection and Analysis*

Data was collected primarily using interviews with respondents. At each organization, interviews were conducted with professionals who had worked directly with GenAI and understood how it was used. Although we spoke with the managers or owners who ran the organization to learn about the company, for information on GenAI use we relied on those who used GenAI regularly. Given the different ways in which GenAI was impacting their work, we focused on a few use cases and in-depth information about those.

Overall, across the three organizations, around six hours of interviews were conducted. Some participants were interviewed multiple times to learn more or to get clarifications regarding their responses. Total interview transcripts contained about 32,000 words. Although this is a small sample size, given the novelty of the technology and the preliminary nature of this work, we believe we were able to reach enough saturation to present important findings.

**Interview Protocol** The interview protocol was designed based on guidelines from the ethnographically-informed interview advanced by Spradley [34]. This approach advocates for having a semi-structured interview protocol and for integrating questions such as the "grand tour" which ask participants to describe a day in their work life. Subsequent questions use their response as a starting point for a more focused approach. The interview protocol then asked participants about specific projects they were working on and the activities they were undertaking. The other design element was a group approach so that participants could bounce off each other and follow-up with more details so that nuances of work practices were captured. Finally, consistent with initial research questions, follow-up interview prompts focused on their learning process and their perspective on newcomer training and future workforce development.

**Analysis**: The data were analyzed iteratively and interpretively in continuous consultation with the literature and relevant prior work [35]. That is, we had read prior work in the area and that formed the basis for our research questions and consequently for the interpretation of data. Our research design and approach, including interview questions, also drew on prior work to help us focus specifically on learning more about the augmentation aspect of GenAI. The interview transcripts were read and themes identified. Authors then worked on revising the themes into specific areas and then further refinement was undertaken to focus on augmentation aspects and other important concerns brought up by the participants.

## IV. Findings

In this section we present findings from the field study. For each research question, we presenting findings across each of the research site and then present an analysis or comparison (see Figure 2).

### A. Human-GenAI Augmentation

**Project Concept**: For this project, the GenAI augmentation was at a more technical level and participants used GenAI in three main ways. First, integrating GenAI in existing solutions that already used some form of machine learning, for instance, ChatBots. Second, creating customized GenAI solutions for clients that work on native systems to overcome the problem of lack of privacy when using publicly available tools or even subscription-based models. Third, generating new content, specifically images, for projects that were user-interface focused.

From the human-GenAI augmentation perspective, there was augmentation of existing products or solutions using GenAI to make use of technologies like large language models (LLMs). For instance, a ChatBot that worked by using pre-fed responses or a static database could be tweaked to use LLMs to generate new responses and use those and the user queries to provide novel answers. The augmentation for this project, occurred at a highly-technical level. At the company level, they were augmenting their expertise by building knowledge about how GenAI can be used and this was shared with other projects and teams. This allowed them to build and demo even more concepts to clients.

The team also identified several limitations with using GenAI. For instance, one critical barrier was the stability of the model output, especially when the models changed over time. Thus, the team had to keep modifying their solution as the model changed, making the incorporation of GenAI applications difficult for stable products for clients. Another concern by clients was the privacy of their data and how they could not trust if the information they used with a GenAI application would remain private. This also applied for any novel product idea as the idea itself would become part of the GenAI application's training data. Finally, as a mid-level firm the cost of computing time and of prompting was prohibitive for them.

**Project Code**: In this project, GenAI played a more central role in augmenting the work practices of the participant directly. GenAI, in this case Co-Pilot was a component of the integrated development environment (IDE) that the team was using. The junior developers, who had not worked with any of the development languages or technologies before, were provided a detailed design for the overall software and were asked to use GenAI to help them with writing the code. Through the design, it was made sure that the developers knew how all the pieces fit and expert guidance was available to them if needed. Furthermore, GenAI was used to develop a range of test cases and thereby improved the quality of the code. Overall, the ability to query through a natural dialogue was very helpful for the junior developers.

The limitations of this approach identified by the team was that for this augmentation to work, a very detailed a high quality design of the overall software is needed. Otherwise, the team working on the project does not have the expertise to do course correction as they don't have experience with similar projects nor the expertise. There is also an issue of working with very novel tools or applications (e.g., drivers) that might not have made into the GenAI knowledge base. Especially if the new developers are not aware that the information they are using is outdated, they are likely to get stuck.

**Project Content**: The augmentation for the content team ranged quite a bit from use for mundane tasks to brainstorming assistance for new ideas. At the most basic level, they used it to revise text or to tweak images for a campaign. In certain cases, it helped them find examples to use that were a better fit and specific; something that would earlier take them a long time if they simply used a search engine. For instance, for one of the campaigns they needed a an idiom in Hindi (*muhavara*) and they could find one by prompting ChatGPT). They said that once you have figured out what works, easier to get to the end result. Overall, they used GenAI sparingly compared to the other two teams but use it to augment creativity and routine tasks.

The content team recognized several limitations of GenAI augmentation. First, it could assist in being creative but the output of the system itself was quite stale and not usable directly. Their job required them to be different from others and GenAI gave the same output to everyone. Second, the training data used was limited in diversity and for the context in which they worked, the output was often not useful. For instance, if they wanted an image of an Indian person sitting in an office they would not get it. Finally, the copyright and intellectual property concern was very high for them as they did not want to give their novel ideas to the system nor end up using an output that was copied from another source and there was no sure way of knowing this.

| Project | Human-AI Augmentation | Knowledge and Learning | Workforce Concerns & Opportunities |
|---|---|---|---|
| Concept | *Augmentation*: Augmentation of existing products using large language models (LLMs) and augmentation of organizational expertise by building knowledge about how GenAI can be used. *Limitations*: Unstable models, lack of transparency, privacy and intellectual property concerns, and prohibitive cost of using the system. | Participants had a high level of computing expertise and used that to learn about GenAI. Read developer docs and other technical details of how GenAI worked before using it. | Difficult to keep up with new technologies; the high cost of applications makes it hard to learn and develop new skills; some concern with automation. Opportunity to develop new skills faster. |
| Code | *Augmentation*: GenAI integrated with development environment; assisted with generating code and developing test cases. *Limitations*: Requires high quality and detailed design; GenAI often lacks information on current software updates or changes. | Expertise varied and senior team members had more knowledge. Read documentation to learn how to use, shared expertise within the team. | Concern with significant potential for automation; skill erosion and hard to know what to learn and teach. Lowers barrier to entry, opportunity to learn without worrying about syntax. |
| Content | *Augmentation*: Text and image editing, finding examples for a concept, and brainstorming for new ideas. *Limitations*: Lack of novelty, stereotypical results, copyright and intellectual property concerns. | Learned primarily through trial and error; limited to no knowledge of how the GenAI application actually worked. | Easy to become reliant on GenAI without developing core competencies; creative ideas and novelty comes from experience; minimal potential for automation. |

Fig. 2. Findings

*B. GenAI Knowledge and Learning*

**Project Concept**: From a technical viewpoint of what GenAI is and how it works, participants working on this project had a high level of expertise. They had the basic understanding of how GenAI systems are developed and how they function. What kinds of responses can be expected from the systems and what might be the limitations. They understood how to call APIs, fine tune models, and customize GenAI within the context of the projects they were working on. They were also able to work with 'open-source' models that were available and run smaller applications in house.

To develop expertise on the use of GenAI, the participants relied to a large extent on their prior knowledge. They were already using machine learning and data mining on other projects and had knowledge of different algorithms and models, and experience at using them. In their formal studies, they had studied in depth about these topics, especially during graduate studies. They used that prior knowledge to further build their expertise by reading all the documentation that was provided with each release of a GenAI application. They also read research papers that were available and blogs from other users on how to use GenAI.

**Project Code**: The expertise on the team varied and the senior people knew more about how GenAI worked but largely the team knew how to use GenAI as part of their IDE to help with developing software, primarily, writing code. The two junior people knew the basics of programming but not necessarily the language they were developing the code in. They knew how to use an IDE and also how to integrate responses they got. They would earlier search online and get code from an online resource such as a blog or Stack Overflow but through Co-Pilot they could do the same within their IDE.

They learned about how to use the system largely through documentation released with GenAI application They picked up new knowledge of the language they had to use while they worked with it. They relied on information they found online but also reached out to the senior people on the team who could guide them when needed.

**Project Content**: The participants on this project had a varied knowledge of GenAI in terms of what it was or how it worked but for the most part their technical expertise was quite low. Through trial and error they had learned to use the system but they needed to know very little beyond whether they could trust the system. They also needed to learn about the different applications that were available to them, including both standalone as well as new functionality in existing software (e.g. Adobe Photoshop).

They largely learned about GenAI through online sources and through trying out different applications. They also learned about it from peer groups and other friends in the industry who were using it.

*C. GenAI and Future Workforce Concerns and Opportunities*

**Project Concept**: Participants reported that it was hard to keep up with the technical advances in the area of GenAI and to implement new solutions that were actually beneficial to their clients. They also expressed a concern with the high cost of using GenAI and how smaller firms like theirs were getting left out as they could not easily afford the infrastructure to work with GenAI, especially in terms of developing their own versions or models. The participants spoke about the time and effort it took to learn new skills by reading research papers, trying out new releases, etc. They did not see automation being a concern in what they did given all the flux around most applications, but in the future they saw some scope for automating of smaller tasks within their work.

The participants realized that GenAI would become a part of many of their work practices but they cautioned that for newcomers it was important to approach its use cautiously. They still needed to learn the basics of computing if they wanted to work at an advanced technical level. They suggested that novices should use it to learn but not become dependent on it. They should understand how it is generating the code and what might be the limitations of the output. They realized that not everyone will be able to get to the highest level of GenAI use, such as fine tuning a model, but there was still lots to learn before reaching that stage.

**Project Code**: Given the ease with which novice programmers were able to develop code for a new application, the participant expressed that there was significant potential for automation and not just augmentation in the software development context. Already there are signs of this as GenAI applications are able to generate code for full fledged applications that just needs to be modified before use. The participant worried that if novices were not trained on higher level software development tasks, who would be the next software architects.

One positive aspect that the participant saw was that using GenAI for accomplishing tasks could be motivating for novices as many errors such as syntax were not a hurdle to them generating good code or product. The capabilities of what they could do with the help of GenAI gave them quick successes and made them want to do more. Another opportunity was that the augmentation lowered the entry barrier into software development, which indirectly might lead to new uses and developments of technologies that have not yet been foreseen.

**Project Content**: The participants in this company expressed a concern with newcomers becoming too reliant on GenAI and not developing or honing the skills to learn how to be creative or come up with novel ideas. They said that it was important for newcomers to be culturally immersed in the world and understand the nuances of language and arts to produce something different and GenAI was more of a synthesizing technology.

The participants stated that although they expected their work to be impacted, given the need for creativity and producing novel ideas, they saw GenAI as an augmenting tool rather than an automation tool.

## V. DISCUSSION AND IMPLICATIONS

This preliminary study is one of the first comparative field studies of the use of GenAI across work settings and sheds important light on the various ways in which the workforce is using GenAI. It raises important questions about how GenAI fits in with technological work practices and the nature of GenAI education that technology students need to receive. Using an augmentation lens to better understand the integration of GenAI in work practices, the findings demonstrate the many ways in which GenAI is being used, and also its limitations. We found GenAI use in the workforce varies by function and also the expertise of the user. GenAI is augmenting many common practices such as software development, but also seeing unique uses such as a brainstorming partner, augmenting creativity and innovation.

There is a range of uses to which professionals are putting GenAI and depending on the specific area that students will end up working in their needs for training might be different. Those who are going to be involved with the development of such technologies need to learn topics like machine learning and algorithms to a much higher level of competency. For students who might end up working with GenAI to augment their software practices, for instance, the competency required will be less with ML and more with tools that integrate with their development environment. Finally, for those who are not necessarily STEM students but will still use these technologies, a lesser level of technical knowledge is required but they still need to know how the tools work so that they are aware of the limitations and also be able to interpret the outcomes appropriately.

Finally, the study also raises questions about the possibility of automating technological work and its implications for students as well. At the same time, there is often more work when certain processes are automated [36]. Overall, most of the participants were both excited by the opportunities of GenAI but also cautious about its use. Some of them worried about automation of certain aspects of their work but had as yet seen little signs of it.

**Implications for AI awareness and literacy**: Many scholars have recently advanced a range of frameworks and guidelines for advancing AI awareness and literacy, as discussed earlier [8]. This is important given the integration of AI across domains [9], [37]. Drawing on prior work, we rated each team on their level of GenAI literacy (see Table II). This rating is relative to each other, i.e., since we did not use any objective measure we used informants' responses to rate them in comparison to other teams in the sample.

Among participants in Project Concept, there was a high level of GenAI literacy. They not only recognized the use of GenAI, they had a good understanding of how it worked and a relatively high expertise in applying it. They could also evaluate a GenAI system to gauge its applicability in the work they were doing. Finally, although they integrated LLMs in the solution, they did not create anything novel but also customized available solutions. In terms of ethical issues, they were concerned about privacy of information they shared with the system.

Participants in Project Code were highly aware of GenAI applications and had a relatively good understanding of how it worked. They were adapt at using them and applying it for the task at hand. They could evaluate the applications to some level, but only as end-users. They had no expertise in creating any components of it. Ethically, they were worried about data privacy when generating code components. They had to do so without sharing client information.

Project Concept participants had a reasonable literacy level when it came to recognizing GenAI applications but not a high level. They had little knowledge of how they worked or were built, but could use them relatively well for their use cases. They could evaluate only through trial-and-error and as end-users. They had no knowledge of how to create them. In terms of ethics, they worried about intellectual property issues, especially the use of copyrighted information.

TABLE II
LITERACY LEVEL OF HUMAN-GENAI AUGMENTATION IN EACH PROJECT

| Construct | Project Concept | Project Code | Project Content |
|---|---|---|---|
| *Recognize* | High | High | Medium |
| *Know and Understand* | High | Medium | Low |
| *Use and Apply* | High | High | Medium |
| *Evaluate* | High | Medium | Low |
| *Create* | Low | None | None |
| *Navigate Ethically* | Data privacy | Data privacy | Intellectual property |

**Implications for student training**: There are a range of augmentation functions that GenAI can play and therefore student training needs to reflect that. They not only need to develop competency in terms of learning how GenAI works, and even how to design GenAI applications and in some cases do fine tuning or even create new GenAI models, they also need to learn different aspects of integration of GenAI with professional practices. Learning to ask questions or prompting is one of the core skills students will have to learn and there are many resources available online. The core concern with student training is an inequity in access to GenAI applications. Given the prohibitive cost of most applications, it is hard for many institutions to afford them for their students and often only students with means to purchase access reap the rewards. There are other ethical considerations as well. GenAI use is highly environmentally unfriendly in terms of energy usage and use of water to cool data centers. Therefore, the environmental impact of training and use also has to be taken into account.

**Implications for faculty development**: Findings from this study also have implications for faculty development. Faculty need to incorporate AI literacy not only through courses on AI but across courses [38], [39]. Yet, the advent of new technologies and their incorporation in the workplace means that faculty usually do not have experience with them. Therefore, a concentrated effort needs to be made to train faculty on the features of GenAI applications and how they can be integrated not only in the workplace, but also in teaching practices. This could also mean redesigning their assignments and assessments to integrate GenAI more effectively. They can also achieve this goal through incorporation of GenAI in their own practices and use this as a teaching moment. In other words, there has to be content about GenAI that has to be taught (literacy levels above) but also the use of GenAI within teaching and learning practices (and this becomes a teachable moment). One way in which this is being accomplished is through assignments that require the use of GenAI and then using the responses as a way to reflect on how GenAI works.

**Limitations**: The primary limitation of this work is the small sample size of the study, especially for the "Code" case, where data was collected from a single individual. Future work needs to broaden both the scope of data collection and the quantity of data that is collected to reach more generalizable conclusions. Furthermore, other forms of data, including participant observations and digital traces can enhance data triangulation and provide more credence to the findings. The study is also limited in its efficacy as the sampling was purposive and data collected was based primarily on convenience of access.

**Areas for Future Research**: Given the novelty of this topic, there is tremendous scope for future work to better understand how GenAI is augmenting knowledge work [40]. Research is also needed to translate the findings of the professional workplace into curriculum and pedagogy for higher education. Given the larger number of GenAI applications that are available and the functions they can perform, what are the essential skills that students need to learn? What is unique about GenAI use that students need to learn during their studies compared to what can they easily learn on the job? Finally, there is evidence that students often have core misconceptions about how GenAI work that hinders their ability to appropriately judge the output of the systems they use [41]. More research is needed to understand how these misconceptions are formed and can be corrected. Future work really needs to dive deeper into the issue of automation versus augmentation given the recent developments in the design and use of AI agents. The relationship of humans with their machines is changing fast in ways that are hard to predict and we needed a timely understanding of what that means for teaching and learning. Finally, although ethical issues were brought up by participants, there was no mention of sustainability, environmental, and social justice concerns and these are critical areas to address when it comes to AI and GenAI use.

## VI. CONCLUSION

In this paper we present a field study of professionals across three companies who use GenAI for augmenting different aspects of their work. We found that there is a wide variation in both how GenAI is used and the knowledge workers in different, but related industries, possess about GenAI. We found a spectrum of use cases and technical fluency among the participants with some projects requiring a high level of technical know-how whereas others relied on a superficial knowledge about GenAI but competency with using it for specific purpose in a project, as needed. We also identify barriers to the use of GenAI not only because of lack of expertise, but also access to tools and developer environments given the high cost of GenAI development. Most participants in our study either used free versions or those already integrated within other applications they used. Finally, participants expressed ethical concerns with the use of GenAI primarily related to privacy of information shared with GenAI applications and issues with copyright and intellectual property that arise from the training dataset. We draw implications for both teaching and learning and argue that it is important to integrated GenAI use in education from an augmentation perspective so that students not only learn about GenAI, but also how it changes educational practices.

ACKNOWLEDGMENT

This work is partly supported by US NSF Awards 2319137, 1954556, and USDA/NIFA Award 2021-67021-35329. Any opinions, findings, and conclusions or recommendations expressed in this material are those of the authors and do not necessarily reflect the views of the funding agencies.

REFERENCES

[1] J. Joskowicz and D. Slomovitz, "Engineers' perspectives on the use of generative artificial intelligence tools in the workplace," *IEEE Engineering Management Review*, 2023.

[2] Microsoft, "2024 work trend index annual report," 2024. [Online]. Available: https://www.microsoft.com/en-us/worklab/work-trend-index/ai-at-work-is-here-now-comes-the-hard-part

[3] A. Johri, A. H. Collier, B. K. Jesiek, R. Korte, and S. C. Brozina, "Workplace learning ecology of software engineers and implications for teaching and learning," in *Proceedings of CSEE&T*, 2024, pp. 1–5.

[4] C. Fleischmann, M. K. Logemann, J. Heidewald, P. Cardon, J. Aritz, and S. Swartz, "Fostering genai literacy in higher education for future workplace preparedness: A mixed-methods study," in *ECIS 2024 Proceedings*, 2024.

[5] A. Yusuf, N. Pervin, and M. Román-González, "Generative ai and the future of higher education: a threat to academic integrity or reformation? evidence from multicultural perspectives," *International Journal of Educational Technology in Higher Education*, vol. 21, no. 1, p. 21, 2024.

[6] N. Ranade, M. Saravia, and A. Johri, "Using rhetorical strategies to design prompts: a human-in-the-loop approach to make ai useful," *AI & SOCIETY*, pp. 1–22, 2024.

[7] H. Subramonyam, R. Pea, C. Pondoc, M. Agrawala, and C. Seifert, "Bridging the gulf of envisioning: Cognitive challenges in prompt based interactions with llms," in *Proceedings of CHI*, 2024, pp. 1–19.

[8] O. Almatrafi, A. Johri, and H. Lee, "A systematic review of ai literacy conceptualization, constructs, and implementation and assessment efforts (2019-2023)," *Computers and Education Open*, p. 100173, 2024.

[9] N. Knoth, M. Decker, M. C. Laupichler, M. Pinski, N. Buchholtz, K. Bata, and B. Schultz, "Developing a holistic ai literacy assessment matrix–bridging generic, domain-specific, and ethical competencies," *Computers and Education Open*, vol. 6, p. 100177, 2024.

[10] E. Brynjolfsson, D. Li, and L. R. Raymond, "Generative ai at work," National Bureau of Economic Research, Tech. Rep., 2023.

[11] R. Raisamo, I. Rakkolainen, P. Majaranta, K. Salminen, J. Rantala, and A. Farooq, "Human augmentation: Past, present and future," *International Journal of Human-Computer Studies*, vol. 131, pp. 131–143, 2019.

[12] S. Paul, L. Yuan, H. K. Jain, L. P. Robert Jr, J. Spohrer, and H. Lifshitz-Assaf, "Intelligence augmentation: Human factors in ai and future of work," *AIS Transactions on Human-Computer Interaction*, vol. 14, no. 3, pp. 426–445, 2022.

[13] S. De'gallier-Rochat, M. Kurpicz-Briki, N. Endrissat, and O. Yatsenko, "Human augmentation, not replacement: A research agenda for ai and robotics in the industry," *Frontiers in Robotics and AI*, vol. 9, p. 997386, 2022.

[14] H. Hassani, E. S. Silva, S. Unger, M. TajMazinani, and S. Mac Feely, "Artificial intelligence (ai) or intelligence augmentation (ia): what is the future?" *AI*, vol. 1, no. 2, p. 8, 2020.

[15] C. Dede, A. Etemadi, and T. Forshaw, "Intelligence augmentation: upskilling humans to complement ai," *The Next Level Lab at the Harvard Graduate School of Education. President and Fellows of Harvard College: Cambridge, MA*, 2021.

[16] L. Zhou, S. Paul, H. Demirkan, L. Yuan, J. Spohrer, M. Zhou, and J. Basu, "Intelligence augmentation: Towards building human-machine symbiotic relationship," *AIS Transactions on Human-Computer Interaction*, vol. 13, no. 2, pp. 243–264, 2021.

[17] J. Trevelyan, "Reconstructing engineering from practice," *Engineering Studies*, vol. 2, no. 3, pp. 175–195, 2010.

[18] ——, "Transitioning to engineering practice," pp. 821–837, 2019.

[19] R. Stevens, A. Johri, and K. O'connor, "Professional engineering work," *Cambridge handbook of engineering education research*, pp. 119–137, 2014.

[20] L. Plowman, Y. Rogers, and M. Ramage, "What are workplace studies for?" in *Proceedings of ECSCW: 10–14 September, 1995, Stockholm, Sweden*. Springer, 1995, pp. 309–324.

[21] E. Hutchins, *Cognition in the Wild*. MIT press, 1995.

[22] A. Johri, "Lifelong and lifewide learning for the perpetual development of expertise in engineering," *European Journal of Engineering Education*, vol. 47, no. 1, pp. 70–84, 2022.

[23] D. Baidoo-Anu and L. O. Ansah, "Education in the era of generative artificial intelligence (ai): Understanding the potential benefits of chatgpt in promoting teaching and learning," *Journal of AI*, vol. 7, no. 1, pp. 52–62, 2023.

[24] N. Ranade and D. Eyman, "Introduction: Composing with generative ai," p. 102834, 2024.

[25] J. Li, "How far can we go with synthetic user experience research?" *Interactions*, vol. 31, no. 3, p. 26–29, may 2024. [Online]. Available: https://doi.org/10.1145/3653682

[26] E. Ras, F. Wild, C. Stahl, and A. Baudet, "Bridging the skills gap of workers in industry 4.0 by human performance augmentation tools: Challenges and roadmap," in *Proceedings of PETRA*, 2017, pp. 428–432.

[27] M. Furendal and K. Jebari, "The future of work: Augmentation or stunting?" *Philosophy & Technology*, vol. 36, no. 2, p. 36, 2023.

[28] L. Tankelevitch, V. Kewenig, A. Simkute, A. E. Scott, A. Sarkar, A. Sellen, and S. Rintel, "The metacognitive demands and opportunities of generative ai," in *Proceedings of the CHI Conference on Human Factors in Computing Systems*, 2024, pp. 1–24.

[29] N. McDonald, A. Johri, A. Ali, and A. Hingle, "Generative artificial intelligence in higher education: Evidence from an analysis of institutional policies and guidelines," *arXiv preprint arXiv:2402.01659*, 2024.

[30] P. Luff, J. Hindmarsh, and C. Heath, *Workplace studies: Recovering work practice and informing system design*. Cambridge university press, 2000.

[31] J. A. Maxwell *et al.*, *Designing a qualitative study*. The SAGE handbook of applied social research methods, 2008, vol. 2.

[32] A. Johri, "Conducting interpretive research in engineering education using qualitative and ethnographic methods," *Cambridge handbook of engineering education research*, pp. 551–570, 2014.

[33] M. L. Markus and A. S. Lee, "Using qualitative, interpretive, and case methods to study information technology," *MIS quarterly*, pp. 35–38, 1999.

[34] J. P. Spradley, *The ethnographic interview*. Waveland Press, 2016.

[35] A. L. Strauss, *Qualitative analysis for social scientists*. Cambridge university press, 1987.

[36] L. Bainbridge, "Ironies of automation," in *Analysis, design and evaluation of man–machine systems*. Elsevier, 1983, pp. 129–135.

[37] J. Schleiss, M. C. Laupichler, T. Raupach, and S. Stober, "Ai course design planning framework: Developing domain-specific ai education courses," *Education Sciences*, vol. 13, no. 9, p. 954, 2023.

[38] J. Schleiss, A. Johri, and S. Stober, "Integrating ai education in disciplinary engineering fields: Towards a system and change perspective," *Proceedings of SEFI*, 2024.

[39] J. Schleiss and A. Johri, "A roles-based competency framework for integrating artificial intelligence (ai) in engineering courses," *Proceedings of SEFI*, 2024.

[40] A. Johri, "Augmented sociomateriality: implications of artificial intelligence for the field of learning technology," *Research in Learning Technology*, vol. 30, 2022.

[41] A. Johri, A. Hingle, and J. Schleiss, "Misconceptions, pragmatism, and value tensions: Evaluating students' understanding and perception of generative ai for education," *IEEE Frontiers in Education*, 2024.